
\magnification=\magstep1
\font\twelvebf=cmbx10 at 12pt
\vsize=8.5truein
\hsize=6.0truein
\vskip 0.25in
\baselineskip=16pt plus 0.2pt minus 0.3pt
\parskip=2pt
\parindent=24pt
\hoffset=0.3in
\footline={\ifnum\pageno<2\hfil\else\hss\tenrm\folio\hss\fi}
\def\H{{\cal H}}
\def\Hi{{\cal H}_i}
\def\frac#1#2{\textstyle{#1\over #2}\displaystyle}
\def\L{\hbox{\it \$}}
\def\mn{{\mu\nu}}
\def\B3{{^3\!B}}
\def\D{{\cal D}}
\def\({{\scriptscriptstyle (}}
\def\){{\scriptscriptstyle )}}
\def\[{{\scriptscriptstyle [}}
\def\]{{\scriptscriptstyle ]}}
{\baselineskip=12pt
 \rightline{IFP--423--UNC}
 \rightline{TAR--009--UNC}
 \rightline{May, 1992}}
\vskip 0.35in
\centerline{\twelvebf Quasilocal energy and conserved charges}
\medskip
\centerline{\twelvebf derived from the gravitational action}
\vskip 0.25in
\centerline{J. David Brown\footnote*{Present address: Departments of Physics
and Mathematics, North Carolina State University, Raleigh, NC 27695--8202}
and James W. York, Jr.}
\medskip
\centerline{\it Institute of Field Physics and}
\centerline{\it Theoretical Astrophysics and Relativity Group}
\centerline{\it Department of Physics and Astronomy}
\centerline{\it The University of North Carolina}
\centerline{\it Chapel Hill, NC 27599-3255}
\bigskip\medskip
\centerline{ABSTRACT}
\smallskip
\noindent
{\baselineskip=12pt The quasilocal energy of gravitational and matter fields in
a spatially bounded region is obtained by employing a Hamilton--Jacobi analysis
of the action functional. First, a surface stress--energy--momentum tensor is
defined by the functional derivative of the action with respect to the
three--metric on ${}^3\!B$, the history of the system's boundary. Energy
density, momentum density, and spatial stress are defined by projecting the
surface stress tensor normally and tangentially to a family of spacelike
two--surfaces that foliate ${}^3\!B$. The integral of the energy density over
such a two--surface $B$ is the quasilocal energy associated with a spacelike
three--surface $\Sigma$ whose intersection with ${}^3\!B$ is the boundary $B$.
The resulting expression for quasilocal energy is given in terms of the total
mean curvature of the spatial boundary $B$ as a surface embedded in $\Sigma$.
The quasilocal energy is also the value of the Hamiltonian that generates unit
magnitude proper time translations on ${}^3\!B$ in the direction orthogonal to
$B$. Conserved charges such as angular momentum are defined using the surface
stress tensor and Killing vector fields on ${}^3\!B$. For spacetimes that are
asymptotically flat in spacelike directions, the quasilocal energy and angular
momentum defined here agree with the results of Arnowitt--Deser--Misner in the
limit that the boundary tends to spatial infinity. For spherically symmetric
spacetimes, it is shown that the quasilocal energy has the correct Newtonian
limit, and includes a negative contribution due to gravitational binding. \par}
\vfill\eject
\centerline{\bf I. INTRODUCTION}
\medskip
Considerable effort has been expended in attempts to define quasilocal energy
in general relativity.[1] Some earlier efforts made use of pseudotensor
methods, which led to coordinate dependent expressions whose geometric
meanings were not clear. Some of the more recent efforts have focused on
constructing from the gravitational Cauchy data  mathematicalexhibit certain
physical properties commonly associated with energy. Although
this approach has led to
some interesting mathematical results, no definitive expression for quasilocal
energy has emerged. In this paper we address the problem of quasilocal energy
from a somewhat different perspective.[2] Rather than postulating a set of
properties for quasilocal energy and then searching for a suitable expression,
we allow the action principle for gravity and matter to dictate the definition
of quasilocal energy and its resulting properties. Because of its intimate
connection to the action and the Hamiltonian, we believe our quasilocal energy
is a natural choice. Furthermore, this quasilocal energy has arisen directly in
the study of thermodynamics for  self--gravitating systems, where it plays the
role of the thermodynamic internal energy that is conjugate to inverse
temperature.[3]

The basic idea for our definition of quasilocal energy is best presented by
considering first an analogy. In nonrelativistic mechanics, the time interval
$T$ between initial and final configurations enters the action as fixed
endpoint data. The classical action $S_{c\ell}$, the action functional
evaluated on a history that solves the classical equations of motion, is an
ordinary function of the time interval and is identified as Hamilton's
principal function.[4] Therefore $S_{c\ell}$ satisfies the Hamilton--Jacobi
equation $H = - \partial S_{c\ell}/\partial T$, which expresses the energy
(Hamiltonian) $H$ of the classical solution as minus the time rate of change
of its action. By a  similar analysis, we shall define the quasilocal energy
for gravitational and matter fields in a spatially bounded region as minus
the time rate of change of the classical action.

We will deal with the physics of a spacetime region $M$ that is topothe product
of a three--space $\Sigma$ and a real line interval. The symbol
$\Sigma$ will be used informally to denote either the family of spacelike
slices that foliate $M$ or a particular leaf of the foliation, depending on
the context in which it is used. The boundary of $\Sigma$ is $B$, which need
not be simply connected. The product of $B$ with the line interval is $\B3$,
an element of the three--boundary of $M$. The endpoints of the line interval
are three--boundary elements denoted $t'$ and $t''$. The situation is depicted
in Fig.~1 in which $B$ is chosen to have the topology of a single two--sphere;
because one spatial dimension is suppressed, $B$ is drawn as a closed curve.
Often we will refer to $\B3$ as a three--boundary, although the complete
three--boundary of $M$ actually consists of the sum of $\B3$, $t'$, and $t''$.

Consider the usual action functional $S^1$ for gravity and matter in which the
three--metric is fixed on $\B3$. This fixed three--metric plays a role
analogous to the fixed time interval in nonrelativistic mechanics in that it
determines the separation in time between initial and final configurations.
The action evaluated on a classical solution is a functional of the boundary
three--metric, and in particular it is a function of the proper time
separation between a typical spacelike slice $B$ and its neighboring slice
within the three--boundary. The quasilocal energy associated with the
spacelike hypersurface $\Sigma$ is defined as minus the variation in the
action with respect to a unit increase in
proper time separation between $B$ and its neighboring
two--surface, as measured orthogonally to $B$. (See Fig.~2.)

The quasilocal energy defined in this way equals the value of the Hamiltonian
that generates unit time translations orthogonal to the boundary two--surface
$B$. This particuquasilocal energy. The advantage in following the procedure
outlined above
comes from considering changes in the classical action due to arbitrary
variations in the boundary three--metric. This more general approach leads to
a tensor that characterizes the stress--energy--momentum content of the
bounded spacetime
region. This stress tensor is a surface tensor defined locally on the
three--boundary $\B3$, and it has a simple relation to the gravitational
momentum that is canonically conjugate to the boundary metric.  Proper energy
density is defined by the projection of the stress tensor normal to the
spacelike surface $B$, while proper momentum density and spatial stress are
(respectively) the normal--tangential and tangential--tangential projections of
the stress tensor. The total quasilocal energy is then simply the integral over
$B$ of the energy density. Also, conserved charges are defined using the
surface stress tensor and Killing vector fields (if any) on the boundary $\B3$.
In particular, angular momentum is the conserved charge associated with a
rotational Killing vector, and is equal to the value of the Hamiltonian that
generates spatial diffeomorphisms along that Killing vector field.

The surface stress tensor is not uniquely defined to the extent that the action
$S^1$ for general relativity and matter is itself ambiguous---arbitrary
functionals $S^0$ of the fixed boundary three--metric can always be subtracted
from the action without affecting the equations of motion. This freedom is
sometimes removed by imposing the condition that the action should vanish for
flat empty spacetime.[5] However, such a criterion cannot be implemented
except with special choices of boundary data for which the boundary
three--metric can be embedded in flat spacetime. Because typical three--metric
in a flat spacetime is not possible, this criterion is
not generally applicable.[2] Instead, we partially remove the ambiguity by
insisting that the energy and momentum densities, and therefore the quasilocal
energy and angular momentum, depend only on the canonical variables defined on
$\Sigma$. This restriction implies that (in particular) the quasilocal energy
is a function on phase space. The remaining freedom in the definition of
quasilocal energy is just the freedom to choose the ``zero" of energy. For
example, the quasilocal energy can be chosen to vanish for a flat slice of
flat spacetime. In that case, when the boundary $B$ is at spatial infinity,
the quasilocal energy agrees with the Arnowitt--Deser--Misner[6] energy at
infinity for asymptotically flat spacetimes.[7]

Our quasilocal energy is determined from the two--boundary geometry and the
total mean curvature of $B$, that is, the integral of the trace of the
extrinsic curvature of $B$ as embedded in $\Sigma$. Therefore, given a
spacetime solution, the quasilocal energy needs for its specification a
spacelike two--surface $B$ and a spacelike normal vector field on $B$. The
surface $B$ and its normal vector field determine a timelike unit normal vector
field, which can be viewed as the four--velocities of observers at $B$ whose
rest frames define space at each point of $B$. The quasilocal energy is the
energy naturally associated with these observers. Typically, a fixed
two--surface $B$ has different values of quasilocal energy associated with
different sets of observers passing through $B$. The quasilocal
energy also varies among different two--surfaces whose timelike unit
normals are contained in a common three--boundary $\B3$. So the quasilocal
energy is observer dependent. On the other hand, conserved charges defined from
the surface stress tensor and Killing vector fields on $\B3$ do not
depend on the observers, in the sense
that they are independent of the slice $B$ within the three--boundary
$\B3$ that is used for their evaluation.

In this paper, the Hamilton--Jacobi type analysis leading to the
stress--energy--momentum tensor, quasilocal energy, and conserved charges is
explicitly carried out for general relativity and matter, with the restriction
that the matter should be non--derivatively coupled to the gravitational
field.  The method we use can be applied to any generally covariant action
that describes spacetime geometry and matter. The sign conventions of
Misner, Thorne, and Wheeler[8] are used throughout, and $\kappa$ denotes
$8\pi$ times Newton's constant. In section 2, we present some notation and a
preliminary discussion of the kinematical relationships needed for describing
general relativity in the presence of spatial boundaries. Section 3 contains
the analysis leading to the stress--energy--momentum tensor. There it is shown
that when the bounded spacetime region is the history of a thin surface layer,
the stress tensor yields the Lanczos--Israel tensor[9] describing the
stress--energy--momentum content of a thin surface layer. In Sec.~4, the
energy density, momentum density, spatial stress, and total quasilocal energy
are defined. Also, the Hamiltonian describing general relativity on a manifold
with boundary is derived. Conserved charges are defined in Sec.~5 , with
special attention given to the description of angular momentum. When the
boundary $B$ is at spatial infinity, the angular momentum agrees with the
definition of Arnowitt--Deser--Misner.[6] Section 6 is devoted to an
exploration of various properties of the quasilocal energy, and includes
explicit calculations for spherically symmetric fluid stars and blaIt is shown
that the quasilocal energy of a spherical star agrees in the
Newtonian limit with the energy deduced from Newtonian gravity. In addition,
the first law of black hole mechanics (thermodynamics) for Schwarzschild black
holes is obtained directly by varying the quasilocal energy. Some of the
mathematical details of our analysis are collected in the Appendix.

\bigskip
\centerline{\bf II. PRELIMINARIES}
\medskip
Our notation is summarized in Table~1. The spacetime metric is  $g_\mn$, and
$n^\mu$ is the outward pointing spacelike unit normal to the three--boundary
$\B3$. The metric and extrinsic curvature of $\B3$ are denoted by $\gamma_\mn$
and $\Theta_\mn$, respectively. These spacetime tensors are defined on $\B3$
only, and satisfy $n^\mu \gamma_\mn = 0$ and $n^\mu\Theta_\mn =0$. In
addition, $\gamma^\mu_\nu$ serves as the projection tensor onto $\B3$.
$\gamma_\mn$ and $\Theta_\mn$ can be viewed alternatively as tensors on $\B3$,
denoted by $\gamma_{ij}$ and $\Theta_{ij}$, where the indices $i$, $j$
refer to coordinates on $\B3$. The boundary momentum is $\pi^{ij}$, and is
conjugate to $\gamma_{ij}$ where canonical conjugacy is defined with respect
to the boundary $\B3$ (see Appendix).

Let $u^\mu$ denote the future pointing timelike unit normal to a family of
spacelike hypersurfaces $\Sigma$ that foliate spacetime. The metric and
extrinsic curvature for $\Sigma$ are given by the spacetime tensors $h_\mn$
and $K_\mn$, respectively, and $h^\mu_\nu$ is the projection tensor onto
$\Sigma$. These tensors also can be viewed as (time dependent) tensors on
$\Sigma$, denoted by $h_{ij}$, $K_{ij}$, and $h^i_j=\delta^i_j$. The momentum
canonically conjugate to the spatial metric $h_{ij}$ is denoted by $P^{ij}$.
Also, the spacetime metrdecomposition,[6]
$$ ds^2 = g_\mn dx^\mu dx^\nu
             = -N^2 dt^2 + h_{ij} (dx^i + V^i dt)(dx^j + V^j dt)
             \ ,\eqno(2.1)$$
where $N$ is the lapse function and $V^i$ is the shift vector.

Observe that lower case latin letters such as $i$, $j$, $k$, $\ell$ refer both
to coordinates on $\B3$ and to coordinates on space $\Sigma$. When used as
tensor indices, the meaning of these latin letters is usually clear from the
nature of the tensor. On occasions in which these two index types must be
distinguished, we will underline the indices corresponding to coordinates on
$\Sigma$; for example, $h_{\underline k\underline\ell}$.

Throughout the analysis we assume that the hypersurface foliation $\Sigma$ is
``orthogonal" to $\B3$, meaning that on the boundary $\B3$ the hypersurface
normal $u^\mu$ and the three--boundary normal $n^\mu$ satisfy
$(u\cdot n)\bigr|_{\B3} =0$.  Thus, $n^\mu$ also can be viewed as a vector
$n^i$ in $\Sigma$ with unit length in the hypersurface metric:
$n^i h_{ij} n^j\bigr|_{\B3} = 1$. So the unit normal in spacetime to the
three--boundary $\B3$ is also the unit normal in $\Sigma$ to the two--boundary
$B$. This restriction simplifies enormously the technical details of our
analysis, and has the following logical basis as well. In the canonical
formalism, the boundary $B$ is specified as a fixed surface in $\Sigma$. The
Hamiltonian must evolve the system in a manner consistent with the presence of
this boundary, and cannot generate transformations that map the canonical
variables across $B$. This means that the component of the shift vector normal
to the boundary must be restricted to vanish,  $V^i n_i\bigr|_{B} = 0$. From a
spacetime point of view, this is the condition that the two--boundary evolves
into a three--surface that contains the unit nhypersurfaces $\Sigma$.
Therefore, $u^\mu$ and $n^\mu$ are
orthogonal on $\B3$.

Because of the restriction $(u\cdot n)\bigr|_{\B3} =0$, the metric on $\B3$
can be decomposed as
$$ \gamma_{ij} dx^i dx^j = -N^2 dt^2 + \sigma_{ab} (dx^a + V^a dt)
   (dx^b + V^b dt) \ ,\eqno(2.2)$$
where $x^a$, $a=1, 2$, are coordinates on $B$ and $\sigma_{ab}$ is the
two--metric on $B$. The extrinsic curvature of $B$ as a surface embedded in
$\Sigma$ is denoted by $k_{ab}$. These tensors can be viewed as spacetime
tensors $\sigma_\mn$ and $k_\mn$, or as tensors on $\Sigma$ or $\B3$ by using
indices $i$, $j$, $k$, $\ell$. Also, $\sigma^\mu_\nu$ is the projection tensor
onto $B$.

\bigskip
\centerline{\bf III. STRESS--ENERGY--MOMENTUM TENSOR}
\medskip
Hamilton--Jacobi theory provides a formal basis for identifying the
stress--energy--momentum as dictated by the action. Before presenting this
analysis, it will be useful to provide a quick review of Hamilton--Jacobi
theory as applied to nonrelativistic mechanics. Start with the action in
canonical form,
$$S^1 = \int dt \biggl( p {dx\over dt} - H^1(x,p,t) \biggr) \ .\eqno(3.1)$$
Now parametrize the system by introducing a coordinate $\lambda$ for the
system path in state space (phase space and time). The action becomes
$$S^1 = \int^{\lambda''}_{\lambda'} d\lambda \bigl[p\dot x - \dot t H^1(x,p,t)
    \bigr]  \ ,\eqno(3.2)$$
where the dot denotes a derivative with respect to $\lambda$. Varying this
action gives
$$\delta S^1 = ({\hbox{terms giving the equations of motion}})+ p\,\delta x
   \bigr|_{\lambda'}^{\lambda''} - H^1\delta t\bigr|_{\lambda'}^{\lambda''}
   \ .\eqno(3.3)$$
Observe that by fixing $x$ and $t$ at the endpoints $\lambda'$ and $\lambda''$,
the endpoint (boundary) terms vanconditions, solutions to the equations of
motion extremize $S^1$.

Now restrict the variations in the action to variations among classical
solutions. In this case, the terms in Eq.~(3.3) giving the equations of motion
vanish, leaving
$$\delta S^1_{c\ell} = p_{c\ell}\, \delta x\bigr|_{\lambda'}^{\lambda''} -
    H^1_{c\ell}\,  \delta t  \bigr|_{\lambda'}^{\lambda''} \eqno(3.4)$$
where ``$c\ell$" denotes evaluation at a classical solution. The
Hamilton--Jacobi equations follow from this expression: the classical momentum
and energy at the final boundary $\lambda''$ are
$$\eqalignno{ p_{c\ell}\bigr|_{\lambda''} &= {\partial S^1_{c\ell}\over\partial
     x''}    \ , &(3.5a)\cr
     H^1_{c\ell}\bigr|_{\lambda''} &= - {\partial S^1_{c\ell}\over
     \partial t''} \ .    &(3.5b)\cr}$$
This latter equation says that for a classical history, the energy
(Hamiltonian) at the boundary $\lambda''$ is minus the change in the classical
action due to a unit increase in the final time $t(\lambda'') = t''$.
(Similarly, variation of the initial boundary time $t(\lambda') = t'$ leads to
the energy at $\lambda'$, but with no minus sign because
positive changes in $t'$ decrease, rather than increase, the time interval).

The action for any system is ambiguous in the sense that arbitrary functions of
the fixed boundary data can be added to the action without changing the
resulting equations of motion. For example, for nonrelativistic mechanics
introduce a subtraction term
$$S^0 = \int_{\lambda'}^{\lambda''} d\lambda {dh(t)\over d\lambda} =
    h(t'')-h(t') \ ,\eqno(3.6)$$
where $h$ is an arbitrary function of $t$. The full action is now defined by
$S = S^1 - S^0$. The subtraction just shifts the value of $S^1$ by a
(boundary--data--dependent) constant, and $S$ has the standard
canonical form with Hajust as in Eq.~(3.3) but with $H^1$ replaced by $H$. The
Hamilton--Jacobi
equation for the energy at $\lambda''$ becomes
$$ H_{c\ell}\bigr|_{\lambda''} = - {\partial S_{c\ell}\over \partial t''}
    \ ,\eqno(3.7)$$
so that different subtraction terms lead to different values of energy.
If a particular physical system allows for a subtraction $S^0$ that gives a
$t$--independent Hamiltonian, such a choice is usually preferred.

For general relativity coupled to matter, consider first the action suitable
for fixation of the metric on the boundary[10]
$$ S^1 = {1\over2\kappa} \int_M d^4x \sqrt{-g}\, \Re  + {1\over\kappa}
   \int_{t'}^{t''} d^3x \sqrt{h}\, K - {1\over\kappa} \int_{\B3} d^3x
    \sqrt{-\gamma}  \,\Theta   + S^{\rm m}  \ ,\eqno(3.8)$$
where $S^{\rm m}$ is the matter action, including a possible cosmological
constant term. $S^1$ is a functional of the four--metric $g_\mn$ and matter
fields on $M$ . The notation $\int^{t''}_{t'} d^3x$ represents an integral
over the three--boundary $t''$ minus an integral over the three--boundary
$t'$. The variation in $S^1$ due to arbitrary variations
in the metric and matter fields is
$$\eqalignno{ \delta S^1 =& \,({\hbox{terms giving the equations of
        motion}}) \cr
   &+ ({\hbox{boundary terms coming from the matter action}}) \cr
   &+\int_{t'}^{t''} d^3x\, P^{ij}\delta h_{ij}
   + \int_{\B3} d^3x \,\pi^{ij}\delta\gamma_{ij}  \ .&(3.9)\cr}$$
Here, $P^{ij}$ denotes the gravitational momentum conjugate to $h_{ij}$, as
defined with respect to the spacelike hypersurfaces $t'$ and $t''$, while
$\pi^{ij}$ is the gravitational momentum conjugate to $\gamma_{ij}$, defined
with respect to the  three--boundary $\B3$. We assume that the matter fields
are minimally coupled to gravity, so the matter action contains no derivatives
(Eqs. (A5) and (A8) of the Appendix) as in vacuum general relativity.

The gravitational and matter fields must be restricted by appropriate boundary
conditions, so that the boundary terms in $\delta S^1$ vanish. This is
required for the action to have well defined functional derivatives that yield
the classical equations of motion, and in turn implies that the action
functional is extremized by solutions to those equations of motion. A natural
set of boundary conditions consists in fixing on the boundaries the fields
whose variations appear in the boundary terms of $\delta S^1$, so that the
variations of those fields indeed vanish. We will adopt such boundary
conditions. For the gravitational variables in particular, the boundary
three--metric $\gamma_{ij}$ is fixed on $\B3$, and the hypersurface metric
$h_{ij}$ is fixed on $t'$ and $t''$. (One alternative to $S^1$ is the action
that differs from $S^1$ by the exclusion of the boundary term involving $K$.
In that case, the term $P^{ij} \delta h_{ij}$ in $\delta S^1$ is replaced
by $-h_{ij}\delta P^{ij}$ and the natural boundary conditions include fixed
$P^{ij}$ at $t'$ and $t''$. Such a change does not affect the definition of
the stress--energy--momentum tensor.)

The ambiguity in $S^1$ is taken into account by subtracting an arbitrary
function of the fixed boundary data. Thus, define the action
$$S = S^1 - S^0 \ ,\eqno(3.10)$$
where $S^0$ is a functional of $\gamma_{ij}$. Of course, $S^0$ can depend on
the initial and final metrics $h'_{ij}= h_{ij}(t')$ and
$h''_{ij} = h_{ij}(t'')$ as well, but for present purposes we find no
advantage in allowing for such generality. The variation in
$S$ just differs from the result in Eq. (3.9) by the term
$$-\delta S^0 = -\int_{\B3} d^3x \, {\delta S^0\over\delta
\delta\gamma_{ij}   \equiv  - \int_{\B3} d^3x \, \pi^{ij}_0
     \delta\gamma_{ij} \ ,\eqno(3.11)$$
where $\pi^{ij}_0$ is defined as the functional derivative of $S^0$. Therefore
$\pi^{ij}_0$ is a function of the metric $\gamma_{ij}$ only.

The classical action $S_{c\ell}$, the action $S$ evaluated at a classical
solution, is a functional of the fixed boundary data  consisting of
$\gamma_{ij}$, $h_{ij}'$, $h_{ij}''$, and matter fields. The dependence of
$S_{c\ell}$ on this boundary data is obtained by restricting the general
variation (3.9--11) to variations among classical solutions, which gives
$$\eqalignno{\delta S_{c\ell} =& \,({\hbox{terms involving variations in the
      matter fields}}) \cr
    &+ \int_{t'}^{t''} d^3x \, P^{ij}_{c\ell} \delta h_{ij} + \int_{\B3}
    d^3x \bigl( \pi^{ij}_{c\ell} -\pi^{ij}_0 \bigr) \delta\gamma_{ij}
     \ .&(3.12)\cr}$$
The analogues of the Hamilton--Jacobi equation (3.5a) are the relationships
$$P^{ij}_{c\ell}\bigr|_{t''}  = {\delta S_{c\ell}\over \delta h_{ij}''}
   \eqno(3.13)$$
for the gravitational momentum at the boundary $t''$, and corresponding
relationships for the matter variables at $t''$. (The notation is slightly
awkward: for nonrelativistic mechanics, $\lambda$ is a coordinate while $t$
and $x$ are dynamical variables; for gravity, $t$ and $x$ are coordinates.)

The analogue of the Hamilton--Jacobi equation (3.7) is more subtle. In the
gravitational action, the three--metric components $\gamma_{ij}$ are the fixed
boundary data that determine the time between spacelike hypersurfaces. Then
the analogue of the boundary data $t''$ from the example of nonrelativistic
mechanics is {\it included\/} in $\gamma_{ij}$; but, of course, the boundary
metric provides more than just information about time. It gives the metrical
distance for all spacetime interCorrespondingly, the simple notion of energy in
nonrelativistic mechanics
becomes generalized to a surface stress--energy--momentum tensor for spacetime
and matter, defined by
$$ \tau^{ij} \equiv {2\over\sqrt{-\gamma}} {\delta S_{c\ell}\over\delta
    \gamma_{ij}} \ .\eqno(3.14)$$
The functional derivative of $S_{c\ell}$ is determined from the variation
Eq. (3.12) to be
$$ {\delta S_{c\ell}\over\delta\gamma_{ij}} = \pi^{ij}_{c\ell} - \pi^{ij}_0
    \ ,\eqno(3.15)$$
so the stress tensor becomes
$$ \tau^{ij}  = {2\over\sqrt{-\gamma}}  \bigl(\pi^{ij}_{c\ell} -
    \pi^{ij}_0\bigr)    \ .\eqno(3.16)$$
It is interesting to note the similarity between definition (3.14) and the
standard definition
$$T^\mn =  {2\over\sqrt{-g}} {\delta S^{m}\over\delta g_\mn} \eqno(3.17)$$
for the matter stress tensor $T^\mn$. It should be emphasized that
$\tau^{ij}$ {\it characterizes the entire system, including contributions
from both the gravitational field and the matter fields\/}.

The Hamilton--Jacobi equation (3.15) and the corresponding matter equations
relate the coordinates and momenta of the gravitational and matter fields as
defined at the boundary $\B3$. These boundary variables satisfy the
constraints of general relativity, as well as gauge constraints associated
with invariances of the matter action. In particular, the boundary momentum
constraint reads
$$0 = -2\D_i\pi^{ij} - {\sqrt{-\gamma}} T^{nj} \ ,\eqno(3.18)$$
which is equivalent to the Einstein equation with one index projected normally
to $\B3$ and the other index projected tangentially to $\B3$. Here, $T^{nj}
\equiv T^\mn n_\mu \gamma_\nu^j$ is the matter stress tensor (3.17), with
indices projected normally and tangentially to $\B3$. The momentum constraint
(3.18) implies that $S_{c\ell}$ depends on the boundary data only to
withindiffeomorphisms of $\B3$. It also implies the relationship
$$\D_i\tau^{ij} = -T^{nj} \eqno(3.19)$$
for the surface stress tensor. This result has a form similar to the equation
of motion $\nabla_\mu T^\mn = 0$ for the matter stress tensor, the key
difference being the appearance of a source term $-T^{nj}$ for the divergence
of $\tau^{ij}$. The consequences of Eq. (3.19) are explored in Sec.~5, where
conserved charges are defined.

It is useful to keep in mind that the boundary $B$ need not be simply
connected. An interesting application arises when, for example, $B$ consists
of two concentric, topologically spherical surfaces, $B_1$ and $B_2$. There
are stress tensors associated with each connected part of the boundary:
$\tau^{ij}_1$ and $\tau^{ij}_2$. Consider the limit in which $B_1$ and $B_2$
coincide, so that the three--geometries on the histories of $B_1$ and $B_2$
are identical. The total stress--energy--momentum
$\tau^{ij}_{\rm\scriptscriptstyle SL}$ of the surface layer is just the sum of
$\tau^{ij}_1$ and $\tau^{ij}_2$, which can be written as the difference
$\tau^{ij}_{\rm\scriptscriptstyle SL} = \tau^{ij}_2 - \tau^{ij}_1$ with the
understanding that $\tau^{ij}_1$ is now computed using the outward normal to
$B_2$. Inserting expression (3.16) for the stress tensors, the terms involving
$\pi^{ij}_0$ cancel and the surface layer stress tensor is given by
$$\tau^{ij}_{\rm\scriptscriptstyle SL} = {2\over\sqrt{-\gamma}}
    \bigl( \pi^{ij}_2 -  \pi^{ij}_1 \bigr)\Bigr|_{c\ell} \ .\eqno(3.20)$$
This equation embodies the results of Lanczos and Israel[9] on junction
conditions in general relativity, which relate the jump in momentum
$\pi^{ij}$ to the {\it matter\/} stress--energy--momentum tensor of the
surface layer. Equation (3.20) actually shows that the jump in momentum gives
the {\it total\/} stress--energy--momentthe assumption that the geometries on
each side of the infinitesimally
thin surface layer coincide, the gravitational contribution to
$\tau^{ij}_{\rm\scriptscriptstyle SL}$ vanishes, and
$\tau^{ij}_{\rm\scriptscriptstyle SL}$ equals the matter stress tensor.
Physically, this result reflects the well known absence
of a local gravitational energy--momentum.[8]

\bigskip
\centerline{\bf IV. ENERGY DENSITY, MOMENTUM DENSITY,}
\centerline{\bf SPATIAL STRESS}
\medskip
{}From the stress--energy--momentum tensor, the proper energy density
$\varepsilon$, proper momentum density $j_a$, and spatial stress $s^{ab}$ are
defined by the normal and tangential projections of $\tau^{ij}$ on a
two--surface $B$:
$$\eqalignno{ \varepsilon &\equiv u_iu_j\tau^{ij} = -{1\over\sqrt{\sigma}}
       {\delta S_{c\ell} \over \delta N} \ ,&(4.1a)\cr
   j_a &\equiv -\sigma_{ai}u_j\tau^{ij} = {1\over\sqrt{\sigma}}
       {\delta S_{c\ell} \over \delta V^a} \ ,&(4.1b)\cr
   s^{ab} &\equiv \sigma_i^{a} \sigma_j^{b} \tau^{ij} = {2\over\sqrt{-\gamma}}
       {\delta S_{c\ell} \over \delta \sigma_{ab}} \ .&(4.1c)\cr}$$
The second equalities in Eqs.~(4.1a-c) follow from definition (3.14) for
$\tau^{ij}$ and the relationships
$$\eqalignno{ {\partial\gamma_{ij}/\partial N} &= -{2} u_iu_j /N\ ,&(4.2a)\cr
   {\partial\gamma_{ij}/\partial V^a} &= -{2} \sigma_{a\(i} u_{j\)} /N
         \ ,&(4.2b)\cr
   {\partial\gamma_{ij}/\partial \sigma_{ab}} &= \sigma_{\(i}^{a}
     \sigma_{j\)}^{b}     \ .&(4.2c)\cr}$$
The quantities $\varepsilon$, $j_a$, and $s^{ab}$ are tensors defined on a
generic two--surface $B$. They represent the energy density, momentum density,
and spatial stress associated with matter and gravitational fields on the
spacelike hypersurfquasilocal energy for $\Sigma$ is given by
$$ E = \int_B d^2x\sqrt{\sigma}\varepsilon
     = -\int_B d^2x {\delta S_{c\ell}\over \delta N} \ .\eqno(4.3)$$
This expression is the closest analogue of the Hamilton--Jacobi equation (3.7),
which gives the energy in nonrelativistic mechanics as minus the change in the
classical action due to a unit change in the boundary time $t''$. Here, the
energy of $\Sigma$ is written as minus the change in $S_{c\ell}$ due to a
uniform, unit increase in the proper time between the boundary surface $B$ and
its neighboring two--surface in $\B3$, as measured normally to $B$. (See
Fig.~2.) The change in the classical action due to such a uniform variation is
expressed in Eq.~(4.3) as the integral over $B$
of the local variation  $\delta S_{c\ell}/\delta N$.

{}From the form (3.16) for the stress tensor, the energy density, momentum
density, and spatial stress are each seen to consist of two terms. The first
term is proportional to projections of the classical gravitational momentum
$\pi^{ij}_{c\ell}$, and the second term is proportional to projections of
$\pi^{ij}_0 \equiv \delta S^0/\delta\gamma_{ij}$. The projections of
$\pi^{ij}_{c\ell}$ can be written in terms of the canonical variables
$h_{ij}$, $P^{ij}$, lapse $N$, and shift $V^i$ by making use of Eq.~(A18) of
the Appendix,  while the projections of $\pi^{ij}_0$ can be written as
functional derivatives of $S^0$ by invoking the relationships (4.2). Using
these results, the proper energy density (4.1a), momentum density (4.1b), and
spatial stress (4.1c) become
$$\eqalignno{ \varepsilon &= {1\over\kappa} k\bigr|_{c\ell} +
       {1\over\sqrt{\sigma}}    {\delta S^0 \over \delta N} \ ,&(4.4a)\cr
   j_a &= -2\bigl( \sigma_{a\underline k} n_{\underline\ell} P^{\underline
     k\underline\e       {\delta S^0 \over \delta V^a} \ ,&(4.4b)\cr
   s^{ab} &= {1\over\kappa} [ k^{ab} + (n\cdot a -k)\sigma^{ab} ]\bigr|_{c\ell}
       - {2\over\sqrt{-\gamma}} {\delta S^0 \over \delta \sigma_{ab}}
     \ .&(4.4c)\cr}$$
The momentum density in Eq. (4.4b) can be viewed as a vector in $\B3$ by
changing the $a$ indices to $i$, or as a vector in $\Sigma$ by changing the
$a$ indices to ${\underline i}$.

Thus far, the subtraction term $S^0$ has been treated as an unspecified
functional of the fixed boundary data $\gamma_{ij}$, which arises from an
inherent ambiguity in the action. We now restrict the form of $S^0$ by
demanding that {\it the energy density $\varepsilon$ and momentum density
$j_a$ of a particular spacelike hypersurface $\Sigma$ should depend only on
the canonical variables $h_{ij}$, $P^{ij}$ defined on $\Sigma$\/}. This
requirement implies that $\varepsilon$ and $j_a$ are functions on phase space.
Observe that no such restriction is placed on the spatial stress: $s^{ab}$ is
interpreted as the flux of the $a$ component of momentum in the $b$ direction,
so $s^{ab}$ should depend on the way the canonical data evolve in time. This
dependence is already clear from the presence of the acceleration $a$ of the
timelike unit normal in the expression (4.4c) for the spatial stress. On the
other hand, the first terms in Eqs. (4.4a-b), those that do not involve $S^0$,
are functions only of the canonical variables.

An obvious choice for a subtraction term $S^0$ that satisfies the above
criterion is simply $S^0 = 0$. More generally, the complete expressions for
energy density and momentum density will be functions only of the canonical
variables if $S^0$ is a linear functional of the lapse $N$ and shift $V^a$ on
the boundary. With such a choice for $S^0$, the funcappearing in Eqs. (4.4a-b)
are functions only of the two--boundary metric
$\sigma_{ab}$, which is the projection of the
hypersurface metric $h_{ij}$ onto the boundary $B$.

Restricted to a linear functional of the lapse and shift,  $S^0$ can be
written as
$$S^0 = -\int_{\B3} d^3x \Bigl[ N\sqrt{\sigma} \bigl(k/\kappa\bigr) \bigr|_0 +
   2  \sqrt{\sigma} V^a \bigl(\sigma_{a\underline k} n_{\underline\ell}
    P^{\underline
    k\underline\ell} /\sqrt{h} \bigr) \bigr|_0 \Bigr] \ .\eqno(4.5)$$
Here, $ k\bigr|_0$ and $\bigl(\sigma_{a\underline k}
n_{\underline\ell} P^{\underline k\underline\ell} /\sqrt{h} \bigr)\bigr|_0$
are arbitrary functions of the two--metric $\sigma_{ab}$. As suggested by
their notations, one method of specifying these functions is to choose a
reference space, that is, a fixed spacelike slice of some fixed spacetime,
and then consider a surface in the slice whose induced two--metric is
$\sigma_{ab}$. If such a two--surface exists, it can be used to evaluate $k$
and $\sigma_{a\underline k} n_{\underline\ell} P^{\underline k\underline\ell}
/\sqrt{h}$, yielding the desired functions of $\sigma_{ab}$. With the
subtraction term (4.5), the energy density and momentum density become
$$\eqalignno{ \varepsilon &=  \bigl( k/\kappa\bigr) \Bigr|^{c\ell}_0
    \ ,&(4.6a)\cr
   j_a &= -2 \bigl( \sigma_{a\underline k} n_{\underline\ell} P^{\underline
    k\underline\ell}   /\sqrt{h} \bigr)\Bigr|^{c\ell}_0  \ ,&(4.6b)\cr }$$
where $\bigr|^{c\ell}_0 $ denotes evaluation for the classical solution minus
evaluation for the chosen reference space. Note that ``evaluation for the
classical solution" is actually evaluation for a particular spacelike
hypersurface in the spacetime; that is, evaluation at a particular point in
phase space. By definition, $\varepsilon$ and $j_a$ vanish for the reference
space,to choose the zero of energy and momentum for the system. Observe that
Eq.~(4.6b) can also be written as a tensor equation in $\Sigma$:
$$j^i = -2\bigl( \sigma^i_{ k} n_{\ell} P^{ k\ell} /\sqrt{h}\bigr)
    \Bigr|^{c\ell}_0   \ ,\eqno(4.7)$$
where we have returned to the practice of omitting the underbars on indices
for tensors in $\Sigma$.

The construction of $ k\bigr|_0$ and $\bigl(\sigma_{a\underline k}
n_{\underline\ell} P^{\underline k\underline\ell} /\sqrt{h} \bigr)\bigr|_0$
described above is sensible only if the two--metric $\sigma_{ab}$ indeed can
be embedded in the reference space, and if the embedding is unique. As a
concrete example, choose a flat three--dimensional slice of flat spacetime as
the reference space. In this case, there are a considerable number of
existence and uniqueness results concerning the embedding of a surface in
$R^3$. For example, it is known that any Riemannian manifold with two--sphere
topology and everywhere positive curvature can be globally immersed in
$R^3$.[11] (An immersion differs from an embedding by allowing for
self--intersection of the surface.) The Cohn--Vossen theorem[11] states that
any compact surface contained in $R^3$ whose curvature is everywhere
nonnegative is unwarpable. (Unwarpable means the surface is uniquely
determined by its two--metric, up to translations or rotations in $R^3$.) From
these results it follows that the functions $ k\bigr|_0$ and
$\bigl(\sigma_{a\underline k} n_{\underline\ell} P^{\underline k\underline\ell}
/\sqrt{h} \bigr)\bigr|_0$ are uniquely determined by the flat reference space,
at least for all positive curvature two--metrics with two--sphere topology.
Note in particular that for a flat slice of flat spacetime,
$\bigl(\sigma_{a\underline k} n_{\underline\ell}
P^{\u$P^{\underline k\underline\ell}$ is identically zero.

Henceforth, we shall adopt the flat reference space subtraction term, assuming
its existence and uniqueness for the two--metrics of interest. The quasilocal
energy (4.3) then becomes
$$E = {1\over\kappa} \int_B d^2x \sqrt{\sigma} \bigl(k-k\bigr|_0\bigr)
     \ ,\eqno(4.8)$$
which is ($1/\kappa$ times) the total mean curvature of $B$ as embedded in
$\Sigma$, minus the total mean curvature of $B$ as embedded in flat space. In
this equation, the superfluous ``$c\ell$" has been dropped. Observe that the
energy of a nonflat slice of flat spacetime is not zero. For spacetimes that
are asymptotically flat in spacelike directions, the energy (4.8) with $B$ at
spatial infinity agrees with the ADM energy.[6] In the more usual expression
for the ADM energy,[6] the flat space subtraction is also present, but it is
hidden in the use of ordinary derivatives acting on the metric tensor
components in the asymptotically flat space with Cartesian coordinates.

Now consider the action (3.10) written in canonical form. For simplicity, we
will omit the matter field (and cosmological constant) contribution
$S^{\rm m}$, although its inclusion is straightforward. Using the space--time
split (A20) of the curvature $\Re$ and the trace $\Theta$ of the
three--boundary extrinsic curvature from Eq.~(A17), the action becomes
$$ S = {1\over2\kappa} \int_M d^4x N\sqrt{h} \bigl[R + K_\mn K^\mn -
   (K)^2 \bigr] - {1\over\kappa} \int_{\B3} d^3x N\sqrt{\sigma}\, k  - S^0
    \ .\eqno(4.9)$$
Next, use expressions (A4) and (A5) for the hypersurface extrinsic curvature
and momentum to obtain
$$N\sqrt{h} \bigl[ K_\mn K^\mn - (K)^2 \bigr] = 2\kappa \bigl[ P^{ij}
   \dot h_{ij} -  2P^{ij} D_i V_j - 2\kappa N G_{ijk\ell} P^{ij}
   P^{k\ell}\bwhere  $G_{ijk\ell} = (h_{ik} h_{j\ell} + h_{i\ell} h_{jk} -
h_{ij}
h_{k\ell} )/(2\sqrt{h})$ is the inverse superspace metric. Inserting this
result into the action (4.9) and using the explicit form (4.5) for $S^0$
gives the action in canonical form,
$$ S = \int_M d^4x \bigl[ P^{ij} \dot h_{ij} - N\H - V^i\Hi \bigr]
   - \int_{\B3} d^3x \sqrt{\sigma} \bigl[ N\varepsilon - V^i j_i\bigr]
   \ ,\eqno(4.11)$$
where the gravitational contributions to the Hamiltonian and momentum
constraints are
$$\eqalignno{ \H &= (2\kappa) G_{ijk\ell} P^{ij} P^{k\ell} - \sqrt{h}
      R/(2\kappa) \ ,&(4.12a)\cr
   \Hi &= -2 D_j P_i^j \ .&(4.12b)\cr}$$
In Eq. (4.11), $\varepsilon$ and $j_i$ denote the energy density (4.6a) and
momentum density (4.7), but with ``$c\ell$" omitted in favor of evaluation at
the generic phase space point $h_{ij}$, $P^{ij}$. From the action (4.11), the
Hamiltonian is explicitly determined to be
$$H = \int_\Sigma d^3x \bigl( N\H + V^i\Hi \bigr) + \int_B d^2x \sqrt{\sigma}
   \bigl(  N\varepsilon - V^i j_i \bigr) \ .\eqno(4.13)$$
The quasilocal energy is seen to equal the value of the Hamiltonian that
generates unit time translations orthogonal to the boundary $B$, that is, the
value of $H$ with $N=1$ and $V^i=0$ on the boundary.

\bigskip
\centerline{\bf V. CHARGE}
\medskip
As mentioned in Sec.~3, the stress--energy--momentum tensor describing a
solution to the equations of motion for gravity and matter will satisfy the
relationship
$$\D_i\tau^{ij} = -T^{nj} \ ,\eqno(5.1)$$
where $T^{nj}\equiv T^{\mu\nu} n_\mu\gamma^j_\nu$. This expression is similar
to the familiar equation of motion for the matter stress tensor, namely
$\nabla_\mu T^\mn =0$, and has a similar interpretation as an approximate
local conservati$\tau^{ij}$ in Eq.~(5.1) contains a source term, $-T^{nj}$. To
interpret Eq.
(5.1) as a conservation law, consider a sufficiently small ``box" $\Delta B$
contained in $B$, over a sufficiently short time $\Delta t$, such that the
timelike unit normal $u^i$ is approximately constant, $\D_i u_j \approx 0$.
Contracting Eq.~(5.1) with $u^i$ and integrating over the spacetime region
$\Delta B\Delta t$ gives the approximate conservation law for the
energy--momentum current density $-u_j\tau^{ij}$, whose components are the
proper energy density $\varepsilon$ and proper momentum density $j_a$. This
conservation law states that the increase in time in the total energy--momentum
contained in $\Delta B$ equals the net energy--momentum that enters $\Delta B$
from within $\B3$, {\it plus\/} a contribution from the source $-T^{nj}$. That
contribution is the {\it matter\/} energy--momentum $-u_jT^{nj}$ that passes
through $\Delta B\Delta t$ as it flows across the boundary $\B3$ into $M$.

The conservation law described above is approximate because typically $u^i$ is
not a Killing vector field on $\B3$, so $\D_{\( i} u_{j\)}$ is not zero.  If
the boundary three--metric $\gamma_{ij}$ does possess an isometry, then global
conserved charges can be defined as follows. Let $\xi^i$ denote a Killing
vector field, $\D_{\( i}\xi_{j\)} = 0$, associated with an isometry of the
boundary three--metric. Contract expression (5.1) with $\xi^i$ and integrate
over $\B3$ to obtain
$$-\int^{t''\cap\B3}_{t'\cap\B3} d^2x\sqrt{\sigma}
    \bigl( u_i\tau^{ij}\xi_j\bigr)
   = -\int_{\B3} d^3x \sqrt{-\gamma} T^{ni} \xi_i  \ .\eqno(5.2)$$
This equation naturally motivates the definition
$$Q_\xi(B) \equiv -\int_B d^2x\sqrt{\sigma}  \bigl( u_i\tau^{ij}\xi_j\bigr)
   \eqno(5.3)$$
for the global ``charge" matter stress tensor serves as its source:
$$Q_\xi(t''\cap\B3) - Q_\xi(t'\cap\B3) = -\int_{\B3} d^3x \sqrt{-\gamma}
    T^{ni} \xi_i   \ .\eqno(5.4)$$
Note that when the surface Killing vector field $\xi_i$ can be extended to a
Killing vector field $\xi_\mu$ throughout $M$, Eq. (5.4) can be written as
$$Q_\xi(t''\cap\B3) - Q_\xi(t'\cap\B3) = -\int^{t''}_{t'} d^3x \sqrt{h}
  (u_\mu T^\mn \xi_\nu)  \ .\eqno(5.5)$$
This result follows from expressing the identity $\int d^4x \sqrt{-g}\,
\nabla_\mu (T^{\mn}\xi_\nu) = 0$ in terms of surface integrals over $\B3$,
$t'$, and $t''$.

For many applications, the source term on the right hand side of Eq.~(5.4)
will vanish either because there is no matter in a neighborhood of the
boundary $\B3$, or more specifically because the component of $T^{ni}$ in the
$\xi_i$ direction vanishes. In this case, Eq.~(5.4) describes the conservation
of charge: because $t'\cap\B3$ and $t''\cap\B3$ are arbitrary surfaces within
$\B3$, $Q_\xi$ is independent of the two--surface $B$ used for its evaluation.
This independence applies to arbitrary spacelike surfaces $B$, not just to the
slices constituting some given foliation of $\B3$. On the other hand, the
total energy $E$, defined by Eq. (4.3), is never conserved in this strong
sense. Although $E$ may have the same value on each slice of a carefully
chosen foliation, this value will generally differ from the energy for other
two--surfaces.

The distinction between the charges $Q_\xi$ and energy $E$ is clarified by
using  definitions (4.1a--b) to write the energy--momentum current density as
$$-u_j\tau^{ij} = \varepsilon u^i + j^i \ .\eqno(5.6)$$
The charge (5.3) then becomes
$$Q_\xi = \int_B d^2x\sqrt{\sigma} (\varepsilon u^i + j^i) \xi_i
     \ .\eqno(5.7)$$
Now consider the situation in which a Kiis timelike, has unit length
($\xi^i\xi_i = -1$), and is also surface forming.
Then $\xi_i$ is the unit normal to a particular foliation of the
three--boundary $\B3$, and on each slice of this foliation, the conditions
$\xi_i = u_i$ and $\xi_i j^i = 0$ hold. Comparing the energy expression Eqs.
(4.3) with the charge (5.7) shows that the energy $E$ associated with such a
slice equals minus the charge $Q_\xi$. For other slices that are not
orthogonal to the Killing vector field $\xi_i$, the associated energy will
generally differ from $-Q_\xi$.

An important example of charge is angular momentum, which is defined whenever
the boundary three--metric admits a rotational symmetry. In this case, denote
the Killing vector field by $\phi^i$ and the charge by $J\equiv Q_\phi$. If
the boundary $B$ used to compute $J$ is chosen to contain the orbits of
$\phi^i$, so $\phi^i$ is tangent to $B$, then according to Eq. (5.7) the
angular momentum can be written as
$$J = \int_B d^2x \sqrt{\sigma}\, j_i\phi^i \ .\eqno(5.8)$$
This expresses the total angular momentum as the integral over a two--surface
$B$, with unit normal orthogonal to $\phi^i$, of the momentum density in the
$\phi^i$ direction. Observe that $J$ is minus the value of the Hamiltonian
(4.13) that generates a rotation along $\phi^i$; that is, minus the value of
the Hamiltonian with $N=0$ and $V^i=\phi^i$ on the boundary. From Eq.~(4.1b),
which defines $j_a$ as a functional derivative, $J$ can be identified as the
change in the classical action due to a ``twist" in the boundary three--metric.
The definition (5.8) for $J$ also agrees with the ADM angular momentum at
infinity for asymptotically flat spacetimes.[6]

According to the previous discussion, any change in angular momentum is
governed by $T^{ni}\phi_i$, which is the flux across thematter momentum in the
$\phi^i$ direction. If $T^{ni}\phi_i$ vanishes, then
the angular momentum is conserved. A related and important property of angular
momentum holds whenever the Killing vector field $\phi^i$ on $\B3$ can be
extended throughout $M$. In particular, choose a slice $\Sigma$ containing the
orbits of $\phi^\mu$, so that $u\cdot\phi=0$ and $\phi^i$ is a Killing vector
field on $\Sigma$. Next, write the momentum density (4.7) as simply
$j^i = -2\sigma^i_k n_\ell P^{k\ell}/\sqrt{h}$, where ``$c\ell$" has been
omitted and we have used the flat reference space for which
$P^{k\ell}\bigr|_0$ vanishes. Then from Eq. (5.8) the angular momentum becomes
$$ J = -2\int_B d^2x \sqrt{\sigma} n_i P^{ij} \phi_j /\sqrt{h}
          = \int_\Sigma d^3x (-2D_iP^{ij})\phi_j \ .\eqno(5.9)$$
The term in parenthesis is just the gravitational contribution (4.12b) to the
momentum constraint, which in general reads
$$0 =  -2D_iP^{ij} - \sqrt{h} T^{u j} \ ,\eqno(5.10)$$
with $T^{uj}\equiv -u_\mu T^{\mn} h_\nu^j$ denoting the proper matter momentum
density. Therefore, the total angular momentum is
$$J = \int_\Sigma d^3x \sqrt{h} T^{uj} \phi_j \ ,\eqno(5.11)$$
the matter momentum density in the $\phi^\mu$ direction, integrated over a
slice $\Sigma$ that respects the spacetime symmetry $\phi^\mu$.

The above results reveal a similarity between angular momentum $J$ and the
total electric charge enclosed by a boundary $B$. With the matter momentum
density in the $\phi^\mu$ direction $T^{uj} \phi_j$ playing the role of
electric charge density, Eq.~(5.11) expresses the total charge $J$ as an
integral over space of the charge density. The surface integral expression
(5.8) for $J$ is then analogous to the integral form of Gauss's law; Gauss's
law expresses the total charge as a surface integral of the (radial) electric
integral of the gravitational field $j_i\phi^i$.

Expression (5.11) for $J$ implies that the total angular momentum of any
axisymmetric, vacuum spacetime region vanishes. This result applies in
particular to the Kerr black hole solution, but deserves further comment in
that case. Recall that the spatial sections $t={\rm constant}$ of the Kerr
geometry, where $t$ is the Boyer--Lindquist stationary time coordinate,
contain the axial Killing vector field. These slices have the topology of a
Wheeler wormhole, $R\times S^2$. Therefore a single surface surrounding the
black hole does not constitute a complete boundary for a region of space
$\Sigma$ contained in a $t={\rm constant}$ slice. That is, as assumed above,
$B$ should consist of two disjoint surfaces at different ``radii", and
expression (5.8) for $J$ includes surface integral contributions from both
surfaces. These contributions cancel, giving $J=0$ in agreement with the
result obtained from Eq.~(5.11). The angular momentum of a Kerr black hole is
more appropriately defined by a single surface integral $\int d^2x
\sqrt{\sigma} j_i\phi^i$ over some topologically spherical surface surrounding
the hole. This is the natural definition of $J$ for, say, a star with spatial
topology $R^3$, so with this definition the angular momentum of stars and
black holes are treated on an equal footing.

\bigskip
\centerline{\bf VI. PROPERTIES OF THE ENERGY}
\medskip
One simple property that the {\it quasilocal\/} energy possesses is additivity.
That is, consider space to consist of two possibly intersecting regions
$\Sigma_1$ and $\Sigma_2$, and assume that $\Sigma_1$, $\Sigma_2$,
$\Sigma_1\cup\Sigma_2$, and $\Sigma_1\cap\Sigma_2$ all haso their energies can
be computed from expression (4.8). It follows that the
energy satisfies
$$E(\Sigma_1\cup\Sigma_2) = E(\Sigma_1) + E(\Sigma_2) -
    E(\Sigma_1\cap\Sigma_2)   \ ,\eqno(6.1)$$
because the contributions from the common boundary of any two adjacent regions
will cancel. As a particular example, let $\Sigma_1$ be topologically a ball
with a two--sphere boundary, and let $\Sigma_2$ be topologically a thick shell
surrounding $\Sigma_1$. In this case $\Sigma_1\cap\Sigma_2$ is empty and the
total energy of the ball $\Sigma_1\cup\Sigma_2$ is just the sum $E(\Sigma_1) +
E(\Sigma_2)$.

In order to gain some intuition for the quasilocal energy, consider the case
of a static, spherically symmetric spacetime
$$ds^2 = -N^2 dt^2 + h^2 dr^2 + r^2 (d\theta^2 + \sin^2\theta d\phi^2)
    \ ,\eqno(6.2)$$
where $N$ and $h$ are functions of $r$ only. Let $\Sigma$ be the interior of
a  $t={\rm constant}$ slice with two--boundary $B$ specified by $r = R =
{\rm constant}$. For simplicity, set Newton's constant to unity, $\kappa =
8\pi$. A straightforward calculation of the extrinsic curvature $k_{ab}$ yields
$$k^\theta_\theta = k^\phi_\phi = -{1\over rh}\biggr|_R \ .\eqno(6.3)$$
The acceleration of the timelike unit normal $u^\mu$ satisfies
$$n\cdot a = {N'\over Nh}\biggr|_R \ ,\eqno(6.4)$$
where $N'$ is the derivative of $N$ with respect to $r$. The subtraction term
Eq.~(4.5) is given by
$$S^0 = {1\over 4\pi} \int dt\,d\theta\,d\phi\, NR\sin\theta \ ,\eqno(6.5)$$
which is obtained by using $k$ from Eq.~(6.3) with $h=1$.

{}From the above results, the proper energy density (4.4a) becomes
$$\varepsilon = {1\over 4\pi} \biggl( {1\over r} - {1\over rh}\biggr)\biggr|_R
    \ ,\eqno(6.6)$$
while the proper momentum density (4.4b) vanishes. The trace of the spatial
stress (4.4c) is given by
$$\sigma_{ab} s^{ab   {2\over  N\sqrt{\sigma}}  \sigma_{ab} {\delta
S^0\over\delta\sigma_{ab}}
   \ ,\eqno(6.7)$$
where the functional derivative of $S^0$ can be obtained from
$${\delta S^0\over\delta R} = {\partial\sigma_{ab}\over\partial R}
   {\delta S^0\over\delta\sigma_{ab}} = {2\over R} \sigma_{ab}
   {\delta S^0\over\delta\sigma_{ab}} \ .\eqno(6.8)$$
Combining these last two equations gives
$$\sigma_{ab} s^{ab} = {1\over 4\pi} \biggl( {N'\over Nh} + {1\over rh} -
   {1\over r}   \biggr)\biggr|_R   \ .\eqno(6.9)$$
Also, the quasilocal energy (4.3) is
$$E = \bigl(r-r/h\bigr)\bigr|_R \ .\eqno(6.10)$$
{}From the discussion in Sec. 5, it follows that the conserved charge
associated
with the timelike static Killing vector field with unit normalization at the
boundary is equal to minus the energy for a $t = {\rm constant}$ slice; that
is, minus the energy computed in Eq. (6.10). Similarly, the vanishing of $j_a$
for the $t = {\rm constant}$ slices shows that the angular momentum (5.8) is
zero.

For a simple isentropic fluid with energy density $\rho(r)$ and pressure
$p(r)$, the Hamiltonian constraint $G^t_t = -8\pi \rho$ implies[8]
$$h = \biggl( 1 - {2m\over r}\biggr)^{-1/2} \ ,\eqno(6.11)$$
where
$$m(r) = 4\pi\int_0^r d\bar r\, \bar r^2 \rho(\bar r) + M \ .\eqno(6.12)$$
Similarly, the Einstein equation $G^r_r = 8\pi p$ gives[8]
$${N'\over N} = {m + 4\pi r^3 p\over r^2 - 2mr} \ .\eqno(6.13)$$
The Schwarzschild black hole solution is obtained by choosing $\rho =p= 0$ and
$m=M$, whereas a fluid star solution with $\rho \neq 0$ must have $M=0$ for
the geometry to be smooth at the origin. In each case, the energy is
$$E = R\biggl[1-\biggl( 1-{2m(R) \over R}\biggr)^{1/2}\biggr] \eqno(6.14)$$
with $m(R)$ defined in Eq. (6.12). Observe that for a compact star or black
hole, $m(R)$ is f$E\to m(\infty)$ in this limit, which is precisely the ADM
energy at
infinity.[12]

The Newtonian approximation for  $E$ consists in assuming $m/R$ to be small,
which yields
$$E \approx m + {m^2\over 2R} \ .\eqno(6.15)$$
In this same approximation the first term, $m(R)$, is just the sum of the
matter energy density plus the Newtonian gravitational potential energy
associated with assembling the ball of fluid by bringing the individual
particles together from infinity.[12]  The second term in Eq. (6.15), namely
$m^2/ 2R$, is just {\it minus\/} the Newtonian gravitational potential energy
associated with building a spherical shell of radius $R$ and mass $m$, by
bringing the individual particles together from infinity. Thus, in the
Newtonian approximation, the energy $E$ has the natural interpretation as the
sum of the matter energy density plus the potential energy associated with
assembling the ball of fluid by bringing the particles together from the
boundary of radius $R$. In this sense, $E$ is the total energy of the system
contained within the boundary, reflecting precisely the energy needed to
create the particles, place them in the system, and arrange them in the final
configuration. Any energy that may be expended or gained in the process of
bringing the particles to the boundary of the system, say, from infinity, is
irrelevant.

A related example is obtained by solving expression (6.14) for $m(R)$, which
yields
$$m(R) = E - {E^2\over 2R} \ .\eqno(6.16)$$
If the boundary $R$ is outside the matter, then $m(R) = m(\infty) = E(\infty)$
is the total energy at infinity. Then using the additivity of the quasilocal
energy, Eq.~(6.16) expresses the energy at infinity as the sum of the energy
$E$ within the radius $R$ and the energy $-E^2/(2R)$ outside the radius $R$.The
energy outside $R$ is negative, and in fact equals the Newtonian
gravitational binding energy associated with building a shell of mass $E$ and
radius $R$. For a charged spherically symmetric distribution of matter,
the corresponding analysis yields
$$E(\infty) = E - {E^2\over 2R} + {Q^2\over 2R} \ ,\eqno(6.17)$$
where $Q$ is the total electric charge. In this case, the energy outside $R$
consists of two contributions, the negative gravitational binding energy
$-E^2/(2R)$ and the positive electrostatic binding energy $+Q^2/(2R)$
associated with building a shell of charge $Q$ and radius $R$.

As a final example, consider the black hole solution $m=M$. If the radius $R$
and mass $M$ are changed in such a solution, the energy (6.14) varies according
to
$$dE = \biggl( 1-{1-M/R\over \sqrt{1-2M/R}}\biggr)dR + {1\over\sqrt{1-2M/R}}dM
        \ .\eqno(6.18)$$
Now define the surface pressure by
$$s\equiv {1\over 2} \sigma_{ab} s^{ab} = {1\over8\pi R} \biggl( {1-M/R\over
      \sqrt{1-2M/R}} - 1 \biggr) \ ,\eqno(6.19)$$
where Eq.~(6.13) with $p=0$ has been inserted into the trace (6.9) of the
spatial stress.  The change in energy now becomes
$$dE = -sd(4\pi R^2) + \bigl(8\pi M\sqrt{1-2M/R}\bigr)^{-1} d(4\pi M^2)
   \ ,\eqno(6.20)$$
which is the first law of mechanics for static, spherically symmetric black
holes. This result also has an interpretation as the first law of
thermodynamics for Schwarzschild black holes.[3,5]  Accordingly, the boundary
surface area $4\pi R^2$ and the surface pressure $s$ are thermodynamically
conjugate variables, and $(4\pi M^2)$ is the Bekenstein--Hawking entropy of
the black hole (with $\hbar$ and Boltzmann's constant set equal to unity).
The quantity $\bigl(8\pi M\sqrt{1-2M/R}\bigr)^{-1}$ is the Hawking black
hole temperature blueshifted from infinity to the finite radius $R$.

\centerline{\bf APPENDIX: KINEMATICS}
\medskip
The spacetime metric is $g_\mn$, and $u^\mu$ is the future pointing timelike
unit normal for a family of spacelike hypersurfaces $\Sigma$ that foliate
spacetime. The normal is proportional to the gradient of a scalar function
$t$ that labels the hypersurfaces,  so that $u_\mu = -N t_{,\mu}$ where $N$ is
the lapse function fixed by the condition $u\cdot u = -1$. A vector field
$T^\mu$ (or tensor field, in an obvious generalization) is said to be
``spatial" (or tangent to $\Sigma$) if it satisfies $u\cdot T =0$. The metric
on  $\Sigma$ is defined as the spatial tensor
$$h_\mn = g_\mn + u_\mu u_\nu \ ,\eqno(A1)$$
which is induced on $\Sigma$ by the spacetime metric $g_\mn$. Note that
$h^\nu_\mu = g^{\nu\sigma} h_{\sigma\mu}$ is the identity for spatial tensors.

The covariant derivative $D_\mu$ compatible with the spatial metric $h_\mn$
that acts on spatial tensors is defined by projecting the spacetime covariant
derivative $\nabla_\mu$. That is, $D_\mu$ is defined by $D_\mu f =
h_\mu^\alpha\nabla_\alpha f$ for any scalar function $f$, $D_\mu T^\nu =
h_\mu^\alpha h^\nu_\beta \nabla_\alpha T^\beta$ for any spatial vector
$T^\mu$, and similarly for higher rank tensors. The extrinsic curvature of
$\Sigma$ as embedded in $M$ is defined by
$$\eqalignno{ K_\mn &= -{1\over2} \L_u h_\mn &(A2.a)\cr
   &= -h_\mu^\alpha \nabla_{\alpha}u_{\nu} \ ,&(A2.b)\cr}$$
where $\L_u$ is the Lie derivative along $u^\mu$. The expression (A2.b) for
$K_\mn$ is symmetric in $\mu$ and $\nu$ because the unit normal $u^\mu$ is
surface forming, and has vanishing vorticity $h^\alpha_{\mu} \nabla_\alpha
u_{\nu} - h^\alpha_{\nu} \nabla_\alpha u_{\mu}= 0$.

It is convenient to introduce coordinates that are adapted to the foliation by
choosing{1, 2, 3}$, lie in the surfaces $\Sigma$ and  $\partial/\partial x^i$
are
spatial vectors. In these adapted coordinates, the normal satisfies $u_\mu =
-N\delta_\mu^0$, and spatial vector fields $T^\mu$ have vanishing
contravariant time components, $T^0 = 0$. From the definition (A1), it
follows that the space components $h^{ij}$ of the contravariant tensor
$h^\mn$ form the matrix inverse of the metric components $h_{ij}$, so that
$h_{ik}h^{kj}=\delta_i^j$. The space components of a spatial vector are raised
and lowered with $h_{ij}$ and its inverse $h^{ij}$, since $T_i \equiv
g_{i\nu}T^\nu = h_{ij} T^j$ and  $T^i \equiv g^{i\nu}T_\nu =h^{ij} T_j$. In
particular, the spacetime tensors $D_\mu f$, $D_\mu T^\nu$, and $K_\mn$ are
tangent to $\Sigma$, so $D_i f$, $D_i T^j$, and $K_{ij}$ are tensors on
$\Sigma$ with indices raised and lowered by the metric $h_{ij}$.

Using the adapted coordinates, the spacetime metric can be written
according to the usual ADM decomposition,[6]
$$\eqalignno{ ds^2 &= g_\mn dx^\mu dx^\nu = (h_\mn - u_\mu u_\nu)
         dx^\mu dx^\nu \cr
        &= -N^2 dt^2 + h_{ij} (dx^i + V^i dt)(dx^j + V^j dt) \ ,&(A3)}$$
where $V^i = h_0^i = -Nu^i$ is the shift vector. Also, the space components of
the extrinsic curvature (A2) become
$$K_{ij} = -{1\over 2N} \bigl[ \dot h_{ij} - 2D_{\(i} V_{j\)} \bigr]
    \ ,\eqno(A4)$$
where the dot in  $\dot h_{ij}$ denotes a derivative with respect to the
coordinate $t$. In addition, define the momentum for the hypersurfaces
$\Sigma$ as
$$P^{ij} = {1\over2\kappa} \sqrt{h} (K h^{ij} - K^{ij}) \ ,\eqno(A5)$$
where $h = \det(h_{ij})$. This definition is appropriate if, as we assume, the
matter fields are minimally coupled to gravity. (That is, the matter action
does not contain derivatives of the metric ten
The intrinsic and extrinsic geometry of the three--boundary $\B3$ are defined
analogously to the above definitions for the family of hypersurfaces
$\Sigma$. However, in this case, the three--boundary $\B3$ is not treated as a
member of a foliation of the spacetime $M$. (The spacetime topology may
prohibit the extension of $\B3$ into a foliation throughout all of $M$.) Let
$n^\mu$ denote the outward pointing spacelike normal to the boundary $\B3$ and
define the metric on $\B3$ by
$$\gamma_\mn = g_\mn - n_\mu n_\nu \ .\eqno(A6)$$
Likewise, define the extrinsic curvature by
$$\Theta_\mn = -\gamma_\mu^\alpha \nabla_{\alpha} n_{\nu} \ .\eqno(A7)$$
Let $\D_\mu$ denote the induced covariant derivative for tensors that are
tangent to $\B3$, defined by projecting the spacetime covariant derivative
onto $\B3$. Introducing intrinsic coordinates $x^i$, $i = {0, 1, 2}$, on the
three--boundary, the intrinsic metric becomes $\gamma_{ij}$. Tensors tangent
to $\B3$, such as the extrinsic curvature (A7), can be written as $\Theta_{ij}$
with indices raised and lowered by $\gamma_{ij}$ and its inverse $\gamma^{ij}$.
Also define the boundary momentum by
$$\pi^{ij} = -{1\over2\kappa} \sqrt{-\gamma} (\Theta\gamma^{ij} - \Theta^{ij})
      \ ,\eqno(A8)$$
where $\gamma = \det(\gamma_{ij})$.

Recall that the hypersurface foliation is restricted by the condition
$(u\cdot n)\bigr|_{\B3} =0$. With this in mind, define the metric on the
two--boundaries $B$, which are the intersections of $\B3$ and the family of
slices $\Sigma$, as
$$\eqalignno{ \sigma_\mn &= h_\mn - n_\mu n_\nu \cr
   &= \gamma_\mn + u_\mu u_\nu \cr
   &= g_\mn + u_\mu u_\nu - n_\mu n_\nu \ .&(A9)\cr}$$
Also define the extrinsic curvature of $B$ as embedded in $\Sigma$ by
$$ k_\mn = - \sigma_\mu^\alpha D_{\alpha} n_{\nu}   \ ,\eqno(A10)$$
where $D_\alpha$ is the covarianare adapted to the foliation $\Sigma$ and the
three--boundary $\B3$,
the line element on $\B3$ is
$$ \gamma_{ij} dx^i dx^j = -N^2 dt^2 + \sigma_{ab} (dx^a + V^a dt)
   (dx^b + V^b dt) \ ,\eqno(A11)$$
where $x^a$, $a=1, 2$, are the coordinates on $B$.
Note that $\sigma_\mn$ and $k_\mn$ are defined only on $\B3$.

We will now outline the steps involved in expressing the extrinsic curvature
of the three--boundary $\B3$ in terms of the intrinsic and extrinsic geometry
of spacetime foliated into spacelike hypersurfaces $\Sigma$. The derivation
makes repeated use of the restriction that on $\B3$ the hypersurface normal
$u^\mu$ and the three--boundary normal $n^\mu$ are orthogonal. With this
condition, $n^\mu$ is a unit normal for both the three--boundary $\B3$
embedded in spacetime $M$, and for the two--boundaries $B$ as embedded in the
hypersurfaces $\Sigma$.

The identity tensor, expressed as $\delta_\mu^\alpha = (h_\mu^\alpha -
u_\mu u^\alpha)$, can be used to split $\Theta_{\alpha\beta}$ into tensors
whose free indices are projected tangentially or normally to the hypersurfaces
$\Sigma$. This results in
$$\eqalignno{ \Theta_\mn = &- h_\mu^\alpha h_\nu^\beta \gamma_\alpha^\rho
  \nabla_{\rho} n_{\beta} - u_\mu u_\nu u^\alpha u^\beta
  \gamma_\alpha^\rho \nabla_{\rho} n_{\beta} \cr
  &+ 2 h_{\(\mu}^\alpha u_{\nu\)}  u^\beta \gamma_\alpha^\rho
  \nabla_{\rho} n_{\beta} \ .&(A12)\cr}$$
Because $u\cdot n =0$ on $\B3$, the projections onto $\B3$ and $\Sigma$
commute, and   the first term on the right hand side of Eq.~(A12) becomes
$$\eqalignno{ - h_\mu^\alpha h_\nu^\beta \gamma_\alpha^\rho \nabla_{\rho}
        n_{\beta}
  &= - \gamma_\mu^\alpha  h_\nu^\beta h_\alpha^\rho \nabla_{\rho} n_{\beta} \cr
   &= -\sigma_\mu^\alpha D_\alpha n_\nu \cr
   &= k_\mn \ .&(A13)\cr}$$
By definitas embedded in $\Sigma$.

For the second term on the right hand side of Eq.~(A12), observe that the
hypersurface normal $u^\mu$ lies in the three--boundary, so that on $\B3$,
$u^\alpha\gamma_\alpha^\rho = u^\rho$.  Also use the relationship $u^\rho
u^\beta \nabla_\rho n_\beta = - u^\rho n_\beta\nabla_\rho u^\beta$, which is
derived by differentiating $u\cdot n\bigr|_{\B3} =0$. Then the second term in
Eq.~(A12) becomes
$$-u_\mu u_\nu u^\alpha u^\beta \gamma_\alpha^\rho
  \nabla_{\rho} n_{\beta} = u_\mu u_\nu n_\beta a^\beta \ ,\eqno(A14)$$
where $a^\beta\equiv u^\rho\nabla_\rho u^\beta$ is the acceleration of the
timelike hypersurface normal $u$.

The third term on the right hand side of Eq.~(A12) is simplified by
recognizing that $\gamma_\alpha^\rho u^\beta \nabla_\rho n_\beta
= - \gamma_\alpha^\rho n^\beta\nabla_\rho u_\beta$, and by using the
relationship $h_\mu^\alpha \gamma_\alpha^\rho = \sigma_\mu^\alpha
h_\alpha^\rho$ on $\B3$. This gives
$$\eqalignno{ 2 h_{\(\mu}^\alpha u_{\nu\)}
    u^\beta\gamma_\alpha^\rho\nabla_\rho n_\beta
   &= -2\sigma_{\(\mu}^\alpha u_{\nu\)} n^\beta h_\alpha^\rho\nabla_\rho
    u_\beta \cr
   &= 2\sigma_{\(\mu}^\alpha u_{\nu\)} n^\beta K_{\alpha\beta} \ ,&(A15)\cr}$$
where $K_{\alpha\beta}$ is the extrinsic
curvature, defined in Eq.~(A2), for the hypersurfaces $\Sigma$.

Collecting these results together, the boundary extrinsic curvature is
expressed as
$$\Theta_\mn = k_\mn +  u_\mu u_\nu n_\beta a^\beta + 2
    \sigma_{\(\mu}^\alpha u_{\nu\)} n^\beta K_{\alpha\beta} \ .\eqno(A16)$$
It immediately follows that the trace of the boundary extrinsic curvature is
$$\Theta = k - n_\beta a^\beta \ .\eqno(A17)$$
Equation (A16) shows that the projection of $\Theta_\mn$ onto $B$ is the
two--boundary extrinsic curvature $k_\mn$, wh$\Theta_\mn$ along the normal
$u^\mu$ is $n_\alpha a^\alpha$. The
``off--diagonal" projection of  $\Theta_\mn$ is given by the ``off--diagonal"
projection of the hypersurface extrinsic curvature $K_\mn$, according to the
relationship $\sigma_\alpha^\mu u^\nu \Theta_\mn = -\sigma_\alpha^\mu n^\nu
K_\mn$. The space--time split of $\Theta_\mn$ also can be written in terms
of the hypersurface momentum (A5) and the boundary momentum (A8) as
$$\pi^{ij} = {N\sqrt{\sigma}\over 2\kappa}\bigl[ k^{ij} + (n\cdot a)
   \sigma^{ij} -k\gamma^{ij}\bigr] - {2N\sqrt{\sigma}\over\sqrt{h}}
   \sigma^{\(i}_{\underline k}
   u^{j\)} P^{\underline k \underline\ell} n_{\underline \ell} \ .\eqno(A18)$$
In this equation, it is necessary to distinguish tensor indices that refer to
coordinates on $\B3$, which are denoted by $i$ and $j$, from tensor indices
that refer to coordinates on the slices $\Sigma$, which are denoted by
$\underline k$ and $\underline \ell$.

The final mathematical ingredient needed for our analysis is the space--time
split of the curvature scalar $\Re$. This is obtained from the decomposition
$$\Re = h^\mn h^{\alpha\beta} \Re_{\mu\alpha\nu\beta} - 2u^\mu u^\nu \Re_\mn
   \ .\eqno(A19)$$
The Gauss--Codazzi relation[8] for the projection of the Riemann tensor onto
$\Sigma$ gives the first term of Eq.~(A19) as  $h^\mn h^{\alpha\beta}
\Re_{\mu\alpha\nu\beta} = R + (K)^2 - K_\mn K^\mn$. The second term of
Eq.~(A19) is rewritten using the Ricci identity $\Re_{\alpha\mu\beta\nu}
u^\nu = 2 \nabla_{\[\alpha} \nabla_{\mu\]} u_\beta$; contracting with
$u^\mu g^{\alpha\beta}$ and rearranging  derivatives gives
$u^\mu u^\nu \Re_\mn = (K)^2 - K_\mn K^\mn + \nabla_\mu (K u^\mu + a^\mu)$.
Together, these results yield
$$\Re = R + K_\mn K^\mn - (K)^2 -2\nabla_\mu (K u^\mu + a^\mu) \ .\eqno(A20)$$

\centerline{\bf ACKNOWLEDGMENTS}
\medskip
We would like to thank R. Gardner, M. Henneaux, T. Jacobson, and M. Kossowski
for valuable insights and discussions. Research support was
received from the National Science Foundation, grant number PHY--8908741.
\bigskip
\centerline{$\underline
      {\qquad\qquad\qquad\qquad\qquad\qquad\qquad\qquad\qquad} $}
\bigskip
\smallskip
\parindent=14pt\frenchspacing

\item{[1]}A sample of the literature on local and quasilocal energy in general
  relativity includes:
A. Komar, Phys. Rev. {\bf 113}, 934 (1959); {\bf 127}, 1411 (1962); {\bf 129},
1873 (1963); (also see C. W. Misner, Phys. Rev. {\bf 130}, 1590 (1963) and J.
Katz, Class. Quantum Grav. {\bf 2}, 423 (1985));
S. W. Hawking, J. Math. Phys. {\bf 9}, 598 (1968);
(also see D. M. Eardley, in {\it Sources of Gravitational Radiation\/}, edited
by L. Smarr (Cambridge University Press, Cambridge, 1979); G. T. Horowitz and
B. G. Schmidt, Proc.  R. Soc. Lond. A {\bf 381}, 215 (1982);
  and D. Christodoulou and S. T. Yau, in {\it Mathematics and General
Relativity\/},  edited by J. Isenberg (American Mathematical Society,
Providence, 1986));
R. Geroch, Ann. NY Acad. Sci. {\bf 224}, 108 (1973);
F. I. Cooperstock and R. S. Sarracino, J. Phys. A: Math. Gen. {\bf 11}, 877
(1978); R. Penrose, Proc. R. Soc. Lond. A {\bf 381}, 53 (1982); in
{\it Asymptotic Behavior of  Mass and Spacetime Geometry\/}, edited by F. J.
Flaherty (Springer--Verlag, Berlin, 1984); (also see K. P. Tod, Proc. R. Soc.
Lond. A {\bf 388}, 457 (1983); Class. Quantum Grav.  {\bf 3}, 1169 (1986);
W. T. Shaw, Proc. R. Soc. Lond. A {\bf 390}, 191 (1983);  R. Penrose and W.
Rindler, {\it Spinors and Space--Time\/}, volume II  (Cambridge Press,
Cambridge, 1986); and L. J. Mason, Class. Quantum Grav.  {\bf 6}, L7
(1989)); M. Ludvigsen and J. A. G. Vickers, J. Phys. A: Math. Gen. {\bf 16},
1155 (1983); D. Lynden--Bell and J. Katz, Mon. Not. R. Astron. Soc. {\bf 213},
21p (1985); M. Dubois--Violette and J. Madore, Commun. Math. Phys. {\bf 108},
213 (1987); J. Katz, D. Lynden--Bell and W. Israel, Class. Quantum Grav.
{\bf 5}, 971 (1988);
R. Bartnik, Phys. Rev. Lett. {\bf 62}, 2346 (1989);
J. Jezierski and J. Kijowski, Gen. Rel. Grav. {\bf 22}, 1283 (1990);
J. Katz and A. Ori, Class. Quantum Grav. {\bf 7}, 787 (1990);
J. M. Nester, Class. Quantum Grav. {\bf 8}, L19 (1991);
G. Bergqvist and M. Ludvigsen, Class. Quantum Grav. {\bf 8}, 697 (1991);
A. J. Dougan and L. J. Mason, Phys. Rev. Lett. {\bf 67}, 2119 (1991).

\item{[2]}J. D. Brown and J. W. York, {\it Mathematical Aspects of Classical
Field Theory\/} edited by M. J. Gotay, J. E. Marsden, and V. E. Moncrief
(American Mathematical Society, Providence, 1992).

\item{[3]}J. W. York, Phys. Rev. {\bf D33}, 2092 (1986); H. W. Braden, B. F.
Whiting, and J. W. York, Phys. Rev. {\bf D36}, 3614 (1987); B. F. Whiting and
J. W. York, Phys. Rev. Lett. {\bf 61}, 1336 (1988); J. W. York, Physica
{\bf A158}, 425 (1989); J. D. Brown, G. L. Comer, E. A. Martinez, J. Melmed,
B. F. Whiting, and J. W. York, Class. Quantum Grav. {\bf 7}, 1433 (1990);
H. W. Braden, J. D. Brown, B. F. Whiting, and J. W. York, Phys. Rev. {\bf D42},
3376 (1990); J. W. York, in {\it Conceptual Problems of Quantum
Gravity\/}, edited by A. Ashtekar and J. Stachel (Birkh\"auser, Boston, 1991);
J. D. Brown, E. A. Martinez, and J. W. York, Phys. Rev. Lett. {\bf 66}, 2281
(1991); J. D. Brown and J. W. York, ``The microcanonical functional integral.
The gravitational field", submitted to Phys. Rev. {\bf D}.

\item{[4]}C. Lanczos, {\it The Variat(University of Toronto Press, Toronto,
1970).

\item{[5]}G. W. Gibbons and S. W. Hawking, Phys. Rev. {\bf D15}, 2752 (1977).

\item{[6]}R. Arnowitt, S. Deser, and C. W. Misner, in {\it Gravitation: An
Introduction to Current Research\/}, edited by L. Witten (Wiley and Sons, New
York, 1962).

\item{[7]}T. Regge and C. Teitelboim, Ann. Phys. {\bf 88}, 286 (1974).

\item{[8]}C. W. Misner, K. S. Thorne, and J. A. Wheeler, {\it Gravitation\/}
(Freeman, San Francisco, 1973).

\item{[9]}C. Lanczos, Phys. Z. {\bf 23}, 539 (1922); Ann. Phys. (Germany)
{\bf 74}, 518 (1924); W. Israel, Nuovo Cimento {\bf 44B}, 1 (1966); {\bf 48B},
463 (1967).

\item{[10]}J. W. York, Found. Phys. {\bf 16}, 249 (1986).

\item{[11]}See, for example, M. Spivak, {\it A Comprehensive Introduction to
Differential Geometry\/} (Publish or Perish, Boston, 1979), Volume 5.

\item{[12]}See Ref.~8, Box 23.1.

\nonfrenchspacing
\vfill\eject
\topinsert \vskip 6truein
\endinsert
\parshape=2 0truein 6truein .80truein 5.20truein
\noindent Figure 1: Spacetime $M$ with boundary $\partial M$, which consists of
initial and final spacelike hypersurfaces $t'$ and $t''$ and a timelike
three--surface  $\B3$. A generic spacelike slice $\Sigma$ has two--boundary
$B$.
\vfill\eject
\topinsert \vskip 6truein
\endinsert
\parshape=2 0truein 6truein .76truein 5.24truein
\noindent Figure 2: Compare with Fig.~1. Here we depict a unit magnitude
``stretch" of the boundary three--metric, normal to the boundary slice $B$.
The change in the classical action induced by such a variation is identified
using the Hamilton--Jacobi equation as minus the quasilocal energy.
\nopagenumbers
\hoffset=-.5truein
\topinsert \vskip 1in \endinsert
\vbox{\offinterlineskip
\hrule
\halign{\vrule#&\strut\quad#\hfil\quad&\vrule#\hskip2pt&&\vrule#&\strut\quad
   \hfil#\hfil\quad\cr
height2pt&\omit&&&\omit&&\omit&&\omit&&\omit&&\omit&&\omit&\cr
&\hfil&&&\hfil&&covariant&&unit&&intrinsic&&extrinsic&&\hfil&\cr
&\hfil&&&metric&&derivative&&normal&&curvature&&curvature&&momentum&\cr
height2pt&\omit&&&\omit&&\omit&&\omit&&\omit&&\omit&&\omit&\cr
\noalign{\hrule}
height2pt&\omit&&&\omit&&\omit&&\omit&&\omit&&\omit&&\omit&\cr
\noalign{\hrule}
height2pt&\omit&&&\omit&&\omit&&\omit&&\omit&&\omit&&\omit&\cr
&spacetime $M$&&&$g_{\mu\nu}$&&$\nabla_\mu$&&\hfil&&
    $\Re_{\mu\nu\sigma\rho}$&&  \hfil&&\hfil&\cr
height2pt&\omit&&&\omit&&\omit&&\omit&&\omit&&\omit&&\omit&\cr
\noalign{\hrule}
height2pt&\omit&&&\omit&&\omit&&\omit&&\omit&&\omit&&\omit&\cr
&hypersurfaces $\Sigma$&&&\hfil&&\hfil&&
    \hfil&&\hfil&&\hfil&&\hfil&\cr
&embedded in $M$&&&$h_{ij}$&&
    $D_i$&&$u_\mu$&&$R_{ijk\ell}$&&$K_{ij}$&&$P^{ij}$&\cr
height2pt&\omit&&&\omit&&\omit&&\omit&&\omit&&\omit&&\omit&\cr
\noalign{\hrule}
height2pt&\omit&&&\omit&&\omit&&\omit&&\omit&&\omit&&\omit&\cr
&3--boundary $\B3$&&&\hfil&&\hfil&&\hfil&&\hfil&&\hfil&&\hfil&\cr
&embedded in $M$&&&$\gamma_{ij}$&&
   ${\cal D}_i$&&$n_\mu$&&\hfil&&$\Theta_{ij}$&&$\pi^{ij}$&\cr
height2pt&\omit&&&\omit&&\omit&&\omit&&\omit&&\omit&&\omit&\cr
\noalign{\hrule}
height2pt&\omit&&&\omit&&\omit&&\omit&&\omit&&\omit&&\omit&\cr
&2--boundary $B$&&&\hfil&&\hfil&&
    \hfil&&\hfil&&\hfil&&\hfil&\cr
&embedded in $\Sigma$&&&$\sigma_{ab}$&&
    \hfil&&$n_i$&&\hfil&&$k_{ab}$&&\hfil&\cr
height2pt&\omit&&&\omit&&\omit&&\omit&&\omit&&\omit&&\omit&\cr
\noalign{\hrule}
}\hrule}

\medskip
\parshape=2 0truein 6.9truein 1.17truein 5.73truein
{\bf Table 1:} A summary of notation. Some spaces are left blank,
either because they are not applicable, or because they are not
needed. The symbol $\Re$ is used for the Riemann tensor on $M$ and
is not a tensor density. The unit normal for $\B3$ embedded in $M$
is also the unit normal for  $B$ embedded in $\Sigma$ by virtue of
the condition $(u\cdot n)=0$ on $\B3$.
\bye